\documentclass[10pt]{article}
\usepackage{graphicx}
\usepackage{latexsym}
\usepackage{amssymb}
\usepackage{amsfonts}
\usepackage{amsmath}
\usepackage{theorem}
\newtheorem{theorem}{Theorem}

\newtheorem{proposition}{Proposition}
\begin{document}

\title{\bf Critical behavior of random spin systems}
\author{Adriano Barra\footnote{King's College London,
Department of Mathematics, Strand, London WC2R 2LS, United Kingdom, and
Dipartimento di Fisica, Universit\`a di Roma
``La Sapienza'' Piazzale Aldo Moro 2, 00185 Roma, Italy;
{\tt<Adriano.Barra@roma1.infn.it>}},
Luca De Sanctis\footnote{ICTP, Strada Costiera 11, 34014 Trieste, Italy;
{\tt<lde\_sanc@ictp.it>}},
Viola Folli\footnote{Dipartimento di Fisica,
   	      Universit\`a di Roma ``La Sapienza'' - Piazzale Aldo Moro 2, 00185
    	      Roma, Italy;
    {\tt<folli@glass.phys.uniroma1.it>}}}

\def\be{\begin{equation}}
\def\ee{\end{equation}}
\def\bc{\begin{center}}
\def\ec{\end{center}}

\maketitle

\begin{abstract}
  We provide a strategy to find in few elementary calculations the
  critical exponents of the overlaps for dilute spin glasses, in
  absence of external field.  Such a strategy is based on the
  expansion of a suitably perturbed average of the overlaps, which is
  used in the formulation of the free energy as the difference between
  a cavity part and the derivative of the free energy itself,
  considered as a function of the connectivity of the model.  We
  assume the validity of certain reasonable approximations, e.g. that
  higher powers of overlap monomials are of smaller magnitude near the
  critical point, of which we do not provide a rigorous proof.
\end{abstract}

\section{Introduction}

Dilute spin glasses are important because of two reasons at least.
Despite their mean field nature, they share with finite-dimensional
models the fact that each spin interact with a finite number of other
spins. Secondly, they are mathematically equivalent to some random
optimization problems. The stereotypical model of dilute spin glasses
is the Viana-Bray model \cite{vb}, which is equivalent to the Random
X-OR-SAT optimization problem in computer science, and the model we
use as a guiding example here. In the original paper \cite{vb} the
equilibrium of the model was studied, even in the presence of an
external field, but the critical behavior was not investigated.  In
the case of fully connected Gaussian models, the critical exponents
were computed in a recent mathematical study \cite{abds}. Here we use
the techniques developed in \cite{bds2} for finite connectivity spin
glasses to extend the methodology of \cite{abds} to the case of dilute
spin glasses. We compute the critical exponents of the overlaps among
several replicas (whose distributions constitute the order parameter
of the model \cite{lds1}).

\section{Model, notations, previous results}

Given $N$ points and families $\{i_{\nu},j_{\nu},k_{\nu}\}$ of i.i.d
random variables uniformly distributed on these points, the (random)
Hamiltonian of the Viana-Bray model is defined on Ising $N$-spin
configurations $\sigma=(\sigma_1,\ldots,\sigma_N)$ through
$$
H_{N}(\sigma,\alpha)=-\sum_{\nu=1}^{P_{\alpha N}}
J_{\nu}\sigma_{i_{\nu}}\sigma_{j_{\nu}}\ ,
$$
where $P_{\zeta}$ is a Poisson random variable with mean $\zeta$,
$\{J_{\nu}=\pm 1\}$ are i.i.d. symmetric random variables and $\alpha>1/2$
is the connectivity. The expectation with respect to all
the (\emph{quenched}) random variables defined so far will be
denoted by $\mathbb{E}$, while the Gibbs expectation at inverse temperature
$\beta$ with respect to this Hamiltonian will be denoted by $\Omega$,
and depends clearly on $\alpha$ and $\beta$. We also define 
$\langle \cdot \rangle = \mathbb{E}\Omega(\cdot)$. The pressure,
i.e. minus $\beta$ times the free energy, is by definition
$$
A_{N}(\alpha)=\frac1N\mathbb{E}\ln\sum_{\sigma}\exp(-\beta 
H_{N}(\sigma,\alpha))\ .
$$
When we omit the dependence on $N$ we mean to have taken the thermodynamic
limit. The quantities encoding the thermodynamic properties of the model
are the overlaps, which are defined on several configurations (\emph{replicas}) 
$\sigma^{(1)},\ldots,\sigma^{(n)}$ by 
$$
q_{1\cdots n}=\frac1N\sum_{i=1}^{N}\sigma^{(1)}_{i}\cdots\sigma^{(n)}_{i} .
$$
When dealing with several replicas, the Gibbs measure is simply the product
measure, with the same realization of the quenched variables, but
the expectation $\mathbb{E}$ destroys the factorization.
We define $\beta_{c}$ as the inverse temperature such that
$2\alpha\tanh^{2}\beta_{c}=1$.
 
We are going to need the \emph{cavity} function given by 
$$
\psi_{N}(\alpha^{\prime},\alpha)=\mathbb{E}\ln\Omega
\exp\beta\sum_{\nu=1}^{P_{2\alpha^{\prime}}}J^{\prime}_{\nu}
\sigma_{k_{\nu}}
$$
where the quenched variables appearing explicitly in this expression
are independent copies of those in $\Omega$. When the perturbation
$\sum_{\nu=1}^{P_{2
    \alpha^{\prime}t}}J^{\prime}_{\nu}\sigma_{k_{\nu}}$ is added to
the Hamiltonian, the corresponding Boltzmann factor will give place to
Gibbs and quenched expectations denoted by
$\Omega^{\prime}_{t}(\cdot),\langle\cdot\rangle^{\prime}_{t}$, and the
subindex $t$ is simply omitted when $t=1$.  This perturbation encoded
in $\psi$, when $\alpha^{\prime}=\alpha$, is equivalent to the
addition of a new spin to the system (which can be interpreted as a
gauging or spin-flip variable). As a consequence \cite{bds2} gauge (or
simply spin-flip in our case) invariant overlap monomials are those
such that each replica appears an even number of times in them, and
are stochastically stable: their average does not depend on the
perturbation in the thermodynamic limit.  The other overlap monomials
are not invariant nor stochastically stable (the two concepts are
equivalent), but their perturbed average can be expressed in terms of
a power series in $t$, with ($t$-independent) stochastically stable
(or invariant) averaged overlap polynomials as coefficients, in the
thermodynamic limit.  This is done by an iterative use of the
following proposition, proven in \cite{bds2}.
\begin{proposition}\label{derivata}
  Let $\Phi$ be a function of $s$ replicas. Then
  the following cavity streaming equation holds
  \begin{multline}\label{stream}
    \frac{d\langle \Phi \rangle^{\prime}_{t}}{dt} =
    -2\alpha^{\prime}\langle \Phi
    \rangle_t^{\prime} +2 \alpha^{\prime}\mathbb{E}[\Omega_t^{\prime}
    \Phi \{ 1 + J\sum_{a}^{1,s}
    \sigma^{(a)}_{i_{1}}\theta +\\
    \sum_{a < b}^{1,s}
    \sigma^{(a)}_{i_{1}}\sigma^{(b)}_{i_{1}} \theta^2 + 
    J\sum_{a < b < c}^{1,s}
    \sigma^{(a)}_{i_{1}}\sigma^{(b)}_{i_{1}}\sigma^{(c)}_{i_{1}}
    \theta^3 + \cdots \} \{ 1 - s
    J\theta\omega \\ + \frac{s(s+1)}{2!}\theta^2
    \omega^{2} 
    -\frac{s(s+1)(s+2)}{3!}J\theta^3\omega^{3}
    + \cdots \}]\ 
  \end{multline}
  where $\omega=\Omega_t^{\prime}(\sigma_{i_{1}})$,
  $\theta=\tanh\beta$.
\end{proposition}
Consider for simplicity the case of $\Phi=q_{1\cdots 2n}$.
In the right hand side above, consisting of the product of two factors
in which each term brings a new overlap multiplying $\Phi$, 
there is only one spin-flip invariant overlap: $q^{2}_{1\cdots 2n}$.
But for the other terms we can use again the streaming equation,
and each non-invariant overlap will be multiplied by a suitable overlap
so that the number of replicas
appearing an odd number of times decreases (by two). Notice though
that each time we use the streaming equation the corresponding exponent of 
$\alpha^{\prime}$ (eventually taken equal to $\alpha$) increase by one
and so does the order of the monomial. Let us be more explicit
in the case of interest, and we will see that we do not need
any explicit calculation, we only need to observe that
monomial of order three or higher are multiplied by $t^2$ or higher
powers of $t$.

\section{The expansion}
 
In the case of $\Phi=q_{12},q_{1234},\ldots$, the previous
proposition yields, integrating back in $dt$ once the thermodynamic
limit is taken
\begin{eqnarray}
\label{q2}
\langle q_{12}\rangle^{\prime}_{t}&=&\tau^{\prime}t\langle q^{2}_{12}\rangle
-2\tau^{\prime 2}t^{2}\langle q_{12}q_{23}q_{31}\rangle+O(q^{4})\\
{}&\vdots&{}\nonumber\\
\label{q2n}
\langle q_{1\cdots 2n}\rangle^{\prime}_{t}&=&\tau^{\prime}\theta^{2n-2}
t\langle q^{2}_{1\cdots 2n}\rangle
+t^{2}O(q^{3})+\cdots
\end{eqnarray}
where $\tau^{\prime}=2\alpha^{\prime}\theta^{2}$
and we neglected monomials with the products
of at least four overlaps. As an example, we gave the explicit
form of the monomial of order three for $n=2$.

These expansions will be used to expand $\psi$ in terms
of averaged stable overlap monomials.

If we take $t=1$ and let $\beta$ be very close to $\beta_{c}$,
we know \cite{bds2} that we can replace $\langle q_{12}\rangle^{\prime}$ 
by $\langle q_{12}^{2}\rangle$, in the left hand side of (\ref{q2}).
This provides a relation, valid at least sufficiently close
to the critical temperature, between $\langle q^{2}_{12}\rangle$
and $\langle q_{12}q_{23}q_{31}\rangle$, as we neglect
the higher order monomials in (\ref{q2}):
\begin{equation}
\label{q-t}
(\tau-1)\langle q^{2}_{12}\rangle=2\langle q_{12}q_{23}q_{31}\rangle
\end{equation}
Notice incidentally that this relation is compatible with the well
known fact \cite{gt1} that the fluctuations of the rescaled overlap
$Nq_{12}^{2}$ diverge only when $\tau\to1$ (and not at higher
temperatures), being $N\langle q_{12}q_{23}q_{31}\rangle$ small (due
to the central limit theorem) as it is the sum of $N^{3}$ bounded
variables dived by $N^{2}$ instead of $N^{3/2}$.

\section{Orders of magnitude}

In the expansions of the previous section, we need to understand which
terms are small near the critical point.  We know that above the
critical temperature all the overlaps are zero, and that those which
are not zero by symmetry become non-zero below the critical
temperature; therefore we assume that slightly below such a
temperature the overlaps are very small.  More precisely, we know that
for instance
$$
\langle q^{2}_{12}\rangle=\mathbb{E}\Omega^{2}(\sigma_{i_{1}}\sigma_{i_{2}})
$$
is very small, and so is therefore
$\Omega^{2}(\sigma_{i_{1}}\sigma_{i_{2}})$.  This means that for
temperatures sufficiently close to the critical one
$\Omega^{4}(\sigma_{i_{1}}\sigma_{i_{2}})$ is negligible as compared
to $\Omega^{2}(\sigma_{i_{1}}\sigma_{i_{2}})$.  In other words
$\langle q^{2}_{1234}\rangle$ is assumed to be of a smaller order of
magnitude than $\langle q^{2}_{12}\rangle$.  Furthermore, if
$q^{2}_{12}$ is small $q^{4}_{12}$ has to be of an even smaller order
of magnitude.
In fact we reasonably assume that $\langle q^{4}_{12}\rangle=
\mathbb{E}\Omega^{2}(\sigma_{i_{1}}\sigma_{i_{2}}\sigma_{i_{3}}\sigma_{i_{4}})$,
which is of order two in $\Omega$, is of a smaller order than $\langle
q^{2}_{12}\rangle$, which is also of order two in $\Omega$. An explanation 
comes from the self-averaging discussed in \cite{dsf}, which tells us that
$\mathbb{E}\Omega(\sigma_{i_{1}}\sigma_{i_{2}}\sigma_{i_{3}}\sigma_{i_{4}})$
is of the same order as
$\mathbb{E}\Omega(\sigma_{i_{1}}\sigma_{i_{2}})
\Omega(\sigma_{i_{3}}\sigma_{i_{4}})$, which is of order two in
$\Omega$, and hence increasing the number of spins in the expectation
$\Omega$ is basically equivalent to increasing the order in $\Omega$.
This is actually proven in a perturbed system \cite{dsf}, but it is
reasonable to assume that the consequences of self-averaging (not the
self-averaging itself) on the orders of magnitude of the considered
quantities is not lost when the perturbation is removed,
and the monomials we have are the result of the 
streaming equation, in which the measure is perturbed.
 Consistently,
(\ref{q-t}) implies that near the critical point $\langle
q_{12}q_{23}q_{31}\rangle$ is smaller than $\langle
q^{2}_{12}\rangle$, and the two critical exponents differ by one.  All
these observations lead to the following criterion. We define the
degree of an averaged overlap monomial as the sum of the degrees of
each overlap in it, where the degree of an overlap is its exponent
times its number of replicas.  For instance $\langle
q^{2}_{1234}q^{2}_{12}q^{2}_{34}\rangle$ is of order $4\times
2+2\times 2+2\times 2=16$.  The definition we just gave coincides with
the one that can be given in terms of $\Omega$ expectations, provided
one multiplies the exponent of each $\Omega$-expectation by the number
of randomly chosen spins appearing in it.  For example $\langle
q^{2}_{1234}q^{2}_{12}q^{2}_{34}\rangle
=\mathbb{E}\Omega^{2}(\sigma_{i_{1}}\sigma_{i_{2}}\sigma_{i_{3}}\sigma_{i_{4}})
\Omega^{2}(\sigma_{i_{1}}\sigma_{i_{2}}\sigma_{i_{5}}\sigma_{i_{6}})$
is of order $2\times 4+2\times 4=16$. Given an integer $m$, a monomial
of order $2m+2$ will be considered negligible, near the critical point
- where all overlaps are very small, with respect to a monomial of
order $2m$.

\section{The transition}

It is well known that all the overlaps are zero above the critical
temperature $1/\beta_{c}$ where the replica symmetric solution
holds, and that below this temperature the overlap between
two replicas fluctuates and its square become non-zero \cite{gt1}.
As pointed out in \cite{vb}, the use of the
replica trick within a quadratic approximation
can only provide the correct transition for the overlap between two replicas,
while overlaps of more replicas would seem to be zero down to lower
temperatures before starting fluctuating. Moreover 
within that method no information about the critical exponents
was found.
Our method allows to gain information about
the critical exponents of all overlap monomials.
Let us start by showing that there is only one critical
point for all overlap monomials.
By convexity, we have
$$
\langle q^{2}_{1\cdots 2n}\rangle=\mathbb{E}\Omega^{2n}
(\sigma_{i_{1}}\sigma_{i_{2}})\geq(\mathbb{E}\Omega^{2}
(\sigma_{i_{1}}\sigma_{i_{2}}))^{n}=\langle q^{2}_{12}\rangle^{n}
$$ 
so that all overlaps are non-zero whenever $\langle q^{2}_{12}\rangle$
is, i.e. below the critical temperature $1/\beta_{c}$.  As a further
example, a slightly more accurate use of convexity yields immediately
$\langle q^{2}_{1234}\rangle\geq\langle q^{2}_{12}q^{2}_{34}\rangle
\geq\langle q_{12}\rangle^{2}$. This means that the critical exponents
of $q^{2}_{1234}$ and $q^{2}_{12}q^{2}_{34}$ cannot be larger than
twice the critical exponent of $q^{2}_{12}$, but cannot be smaller
than this critical exponent itself either, as $\langle
q^{2}_{1234}\rangle \leq\langle q^{2}_{12}\rangle$.

\section{Critical exponents}

We will now relate the free energy to its derivative and to the cavity function. 
The following theorem 
follows easily from the results of
\cite{lds1}, and here we only sketch the proof, based on 
standard convexity arguments. 
\begin{theorem}\label{theorem} 
In the thermodynamic limit, we have
$$
A(\alpha)=\ln 2 +\psi(\alpha,\alpha)-\alpha A^{\prime}(\alpha)
$$
for all values of $\alpha,\beta$,
where $A^{\prime}$ is the derivative of $A$.
\end{theorem}
\textbf{Sketched Proof}.
It was proven in \cite{lds1} that
\begin{multline}\label{var}
A(\alpha)=\lim_{N}[\mathbb{E}\ln\Omega(\sum_{\sigma_{N+1}}\exp
(\beta\sum_{\nu=1}^{P_{2\alpha}}J^{\prime}_{\nu}
\sigma_{k_{\nu}}\sigma_{N+1}))-\\
\mathbb{E}\ln\Omega(\exp-\beta(H^{\prime}_{N}(\alpha/N)))]
\end{multline}
where the quenched variables in $H^{\prime}$ are independent of those
in $\Omega$, just like for the first term in the right hand side. The
second term in the right hand side is easy to compute, at least in
principle \cite{lds1}, and it is the derivative of $A$ multiplied by
$\alpha$, because
$$
\mathbb{E}\ln\Omega(\exp-\beta(H^{\prime}_{N}(\alpha/N)))=
NA(\alpha(1+1/N)-NA(\alpha)\ .
$$
This leads to the result to prove, as
the gauge invariance of $\Omega$ allows to take out the
sum over $\sigma_{N+1}$ as $\ln 2$, and therefore
the first term in the right hand side of
(\ref{var}) is precisely
$\psi$. $\Box$

It is easy to see that \cite{lds1}
\begin{eqnarray}
\label{derivata1}
&&\partial_{1}\psi_{N}(\alpha^{\prime},\alpha)=2\sum_{n}\frac{\theta^{2n}}{2n}
(1-\langle q_{1\cdots 2n}\rangle^{\prime})\ , \\
\label{derivata2}
&&A^{\prime}(\alpha)=\sum_{n}\frac{\theta^{2n}}{2n}
(1-\langle q^{2}_{1\cdots 2n}\rangle)\ .
\end{eqnarray}
From the theorem we have then
$$
A^{\prime}(\alpha)=\partial_{1}\psi(\alpha,\alpha)+
\partial_{2}\psi(\alpha,\alpha)-A^{\prime}(\alpha)-
\alpha A^{\prime\prime}(\alpha)\ .
$$
But we know \cite{bds2} that near the critical point {\em saturation}
$\langle q_{2n}\rangle^{\prime}\to\langle q^{2}_{2n}\rangle$ occurs
in the thermodynamic limit,
so that $\partial_{1}\psi(\alpha,\alpha)\to A^{\prime}(\alpha)$
and therefore we have just proven the next
\begin{proposition} In the thermodynamic limit
\begin{equation}
\label{main}
\partial_{2}\psi(\alpha,\alpha)-\alpha A^{\prime\prime}(\alpha)=0\ .
\end{equation}
\end{proposition}

Notice that if in the statement of Theorem \ref{theorem} we assumed
saturation
$\langle q_{1\cdots 2n} \rangle^{\prime}_{t}\to \langle q^{2}_{1\cdots
  2n}\rangle$ not just for $t=1$ but for all $t$ (once
$\psi(t\alpha,\alpha)$ is written using (\ref{derivata1}) as the
integral of its derivative with respect to $t$), we would obtain
$\psi=2A^{\prime}$ and
$$
A(\alpha)=\alpha A^{\prime}(\alpha)+\ln 2\ ,
$$
which, as the initial condition is easily checked to be 
$A^{\prime}(0)=\ln\cosh\beta$,
gives the well known replica symmetric solution $A(\alpha)=
\ln 2 + \alpha\ln\cosh\beta$. This means
that stability and saturation of the overlaps are
equivalent to replica symmetry.

Now let us analyze (\ref{main}).
We consider $\psi(\alpha^{\prime},\alpha)$ as the integral of its derivative
with respect to its first argument. The derivative, given in (\ref{derivata1}),
contains the perturbed averaged overlaps, which we expand 
using (\ref{q2})-(\ref{q2n}) etc.. In this expansions the 
variable $\alpha^{\prime}$ appears only explicitly in front of 
the averaged overlap monomials, which do not depend on 
$\alpha^{\prime}$, they only depend on $\alpha$.
Therefore we can perform explicitly the integration of these simple
power series in $\alpha^{\prime}$.
The dependence on $\alpha$ of $\psi(\alpha^{\prime},\alpha)$ 
is hence only in the averaged overlap monomials,
and the same holds for $A^{\prime}(\alpha)$, because of
(\ref{derivata2}). 
Therefore the derivatives of $\psi(\alpha^{\prime},\alpha)$ and 
of $A^{\prime}(\alpha)$ with respect to $\alpha$
in (\ref{main}) involve only the averaged overlap monomials.
In other words if we define $\tilde{A}(\alpha^{\prime},\alpha)
=\ln 2+\psi(\alpha^{\prime},\alpha)-\alpha^{\prime}A^{\prime}(\alpha)$,
so that $A(\alpha)=\tilde{A}(\alpha,\alpha)$ thanks
to Theorem \ref{theorem}, equation (\ref{main})
amounts to say that $\partial_{2}\tilde{A}(\alpha,\alpha)=0$.
But since the second argument appears only in the averaged overlap
monomials, we can consider 
$A(\alpha)=\tilde{A}(\alpha,\alpha)\equiv
\hat{A}(\alpha,p_{1}(\alpha),p_{2}(\alpha),\ldots)$
a function of the averaged overlap monomials,
here called $p_{1}(\alpha),p_{2}(\alpha),\ldots$, such that
\begin{equation}
\label{differential}
\partial_{2}\tilde{A}=\sum_{m}\frac{\partial\hat{A}}{\partial p_{m}}
\frac{d p_{m}}{d\alpha}=0\ .
\end{equation}
We can now use (\ref{q2})-(\ref{q2n}) etc. to have an explicit 
expansion of $A(\alpha)$ and deal with the
differential equation (\ref{differential}).
The result is easy to obtain and reads
\begin{multline}\label{aaa}
A(\alpha) = \ln 2  
+\frac{\tau}{2} -\frac{\tau}{4}(\tau-1) \langle q_{12}^2 \rangle + 
\frac{\tau^3}{3} \langle q_{12}q_{23}q_{13} \rangle +
 O(q^4) \\  
 +\theta^2 (\frac{\tau}{4} -
 \frac{\tau}{8}(\tau\theta^{2}-1) \langle q_{1234}^2 \rangle - 
 \frac{3\tau^3}{4} \langle q_{1234}q_{12}q_{34} \rangle 
 + O(q^4))  \\ +O(\theta^4)\ .
\end{multline}
Notice that this expansion extends the one found in \cite{vb}.
As a first approximation we may consider
$$
A(\alpha) \sim \ln 2  
+\frac{\tau}{2} -\frac{\tau}{4}(\tau-1) \langle q_{12}^2 \rangle + 
\frac{\tau^3}{3} \langle q_{12}q_{23}q_{13} \rangle 
$$
and (\ref{differential}) becomes
\begin{equation}
\label{diff1}
-\frac14(\tau-1)\frac{d\langle q^{2}_{12}\rangle}{\alpha}
+\frac13\frac{d\langle q_{12}q_{23}q_{31}\rangle}{\alpha}=0
\end{equation}
because
\begin{eqnarray}
\frac{\hat{A}}{\partial\langle q^{2}_{12}\rangle} 
& = & -\frac{\tau}{4}(\tau-1)
\sim -\frac{1}{4}(\tau-1)\ , \nonumber\\
\frac{\hat{A}}{\partial\langle q_{12}q_{23}q_{31}\rangle} 
& = & \frac{\tau^3}{3}\sim \frac13\ . \nonumber 
\end{eqnarray}
But now the use of (\ref{q-t}) in (\ref{diff1}) offers
$$
-\frac14(\tau-1)\frac{d\langle q^{2}_{12}\rangle}{d\alpha}
+\frac13\frac12\frac{d(\tau-1)\langle q^{2}_{12}\rangle}{d\alpha}=0
$$
from which, after a couple of elementary steps
$$
(\tau-1)\frac{d\langle q^{2}_{12}\rangle}{d(\tau-1)}
-2\langle q^{2}_{12}\rangle=0\ .
$$
This equation is as accurate as close the temperature
is to the critical one, and the solution is easy to find:
$$
\langle q^{2}_{12}\rangle=(\tau-1)^{2}\ ,
$$
describing the critical behavior of the overlap slightly
below the critical temperature. The critical exponent
is hence two.

Notice that (\ref{q-t}) imply 
that $\langle q_{12}q_{23}q_{31}\rangle$
is zero above the temperature $1/\beta_{2}$ and positive slightly
below. Moreover,
(\ref{q-t}) gives the critical exponent for 
$\langle q_{12}q_{23}q_{31}\rangle$:
three.

From our analysis in the previous sections, we 
conclude that the critical exponent of $q^{2}_{1234}$
is strictly larger than three, but no larger than four.
The criterion explained in the section on the
order of magnitudes, together with $\ref{q-t}$
and the critical exponent of $q_{12}^{2}$,
provides a relation between the degree of an 
overlap monomial and its critical  exponent:
degree $2m$ corresponds to critical exponent
$m$. So for instance the critical exponent
of $q_{1\cdots 2n}^{2}$, which is of order $4n$,
is $2n$.

In the infinite connectivity limit
we recover the all the critical exponents for
the fully connected Gaussian SK model \cite{abds}.

\textbf{Remark}.
If we extended the use of
$\langle q_{1\cdots 2n}\rangle^{\prime}\to\langle q^{2}_{1\cdots 2n}\rangle$
to lower temperatures,
such that $2\alpha\theta^{2n}\equiv\tau_{2n}\sim 1$,
we would obtain for $q^{2}_{1\cdots 2n}$,
for all $n$, the same identical differential equation we got
for $q^{2}_{12}$. 
We would then get the same approximated behavior
one gets using the replica method in a quadratic
approximation \cite{vb}: $q^{2}_{2n}$ would be zero
above the temperature such that $\tau_{2n}=1$, then it starts
fluctuating, with critical exponent two. In this sense the 
replica method with quadratic approximation is
equivalent to extending stochastic stability below
the critical point.

\section{Summary and conclusions}

Our strategy requiree the expansion of the
averaged overlaps in powers of a perturbing parameter with
stochastically stable overlap monomials as coefficient (similarly to
the expansion exhibited in \cite{barra} for Gaussian models). This
allowed to write the free energy in terms of overlap fluctuations and
to discover that it does not depend on a certain family of these
monomials. As a consequence, we obtained a differential equation whose
solution, once all small terms are neglected, gave the critical
behavior of the overlaps.

Our method is ultimately based on stochastic stability,
but such a stability is proven or at least believed to hold
in several contexts, therefore generalizations of our method
to finite dimensional spin glasses, to the traveling salesman
problem, to the K-SAT problem, to neural networks 
and to other cases are not
to be excluded and are being studied. We plan on reporting 
soon on these topics \cite{bds5}.

\section{acknowledgments}
The authors are extremely grateful to Peter Sollich for precious
suggestions.
AB is partially supported by the MIUR within the Smart-Life Project (Ministry 
Decree 13/03/2007 n 368) and by Calabria Region -Technological 
Voucher Contract n.11606 of 15/03/2007.



\addcontentsline{toc}{chapter}{References}
\begin{thebibliography}{9}


\bibitem{abds} A. Agostini, A. Barra, L. De Sanctis, 
J. Stat. Mech. {\bf P11015} (2006).

\bibitem{barra} A. Barra, J. Stat. Phys.{\bf 123} (2006). 

\bibitem{bds2} A. Barra, L. De Sanctis, J. Stat. Mech., {\bf P08025} (2007).

\bibitem{bds5} A. Barra, L. De Sanctis, in preparation.
 
\bibitem{lds1} L. De Sanctis, J. Stat. Phys. \textbf{117} (2004).

\bibitem{dsf} L. De Sanctis, S. Franz, ArXiv:mat-ph/0705.2978.

\bibitem{gt1} F. Guerra, F.L. Toninelli, J. Stat. Phys. {\bf 115} (2004).
  
\bibitem{mpv} M. M\'ezard, G. Parisi and M. A. Virasoro, {\em Spin glass theory
and beyond}, World Scientific, Singapore (1987).

\bibitem{vb} L. Viana, A.J. Bray, J. Phys. C {\bf 18} (1985).
  

\end{thebibliography}
\end{document}